\documentclass{ws-p8-50x6-00}

\begin{document}

\title{A Simple Model for the BFKL-DGLAP Transition}

\author{G\"osta Gustafson}

\address{Dept. of Theor. Physics, Lund University, S\"olvegatan 14A, 22362 Lund, Sweden\\E-mail: gosta@thep.lu.se}


\maketitle

\abstracts{A simple model for the gluon structure function at small $x$ is presented, which gives an intuitive picture of the transition between the DGLAP and BFKL regimes.}

\section{Introduction}

When $k_\perp$ is large and $1/x$ is limited we are in the DGLAP regime, and
when $1/x$ is large and $k_\perp$ is limited we are in the BFKL regime. An essential question is then: What is large? Where is the boundary between the regimes, and what is the behaviour in the transition region? We will here present a simple interpolating model, which can illuminate these questions.
The model is relevant for small $x$, and has a smooth transition between large $k_\perp$, where ordered chains dominate, and small $k_\perp$, where non-ordered chains are most important.
We will first discuss the case of a fixed coupling, and after that the case with a running coupling. The results are obtained in collaboration with Gabriela Miu, and a more extensive presentation is given in ref. 1.

\section{Fixed coupling}

For large values of $k_\perp$ (in the ``DGLAP region'') the non-integrated structure function ${\cal F}(x,k_\perp^2)$ is dominated by contributions from chains which are ordered in $ k_\perp$.
 For small $x$ the result is a product of factors $ \frac{3\alpha_s}
{\pi} \cdot \frac{dx_i }{ x_i} \cdot \frac{d k_{\perp,i}^2 }{ k_{\perp,i}^2}$ \cite{DGLAP}:

\begin{eqnarray}
G \equiv \frac{1}{\bar{\alpha}} {\cal F}(x,k_\perp^2) \sim \sum_N \int \prod^N \bar{\alpha}\,\, dl_i \theta(l_{i+1} -l_i)\,\, d\kappa_i \theta(\kappa_{i+1} - \kappa_i) \nonumber \\
{\mathrm {where}} \,\,\,\, \bar{\alpha} \equiv \frac{3\alpha_s}{\pi}, \,\,\,\,l \equiv \ln(1/x)\,\,\,\,{\mathrm {and}} \,\,\,\,\kappa \equiv ln(k_\perp^2)
\label{dglap}
\end{eqnarray}
    
Integration over $\kappa_i$ with the $\theta$-functions gives the phase space for $N$ ordered values $\kappa_i$. The result is $\kappa^N/N!$. The integrations over $l_i$ give a similar result, and thus we obtain the wellknown double log result
\begin{equation}
\frac{1}{\bar{\alpha}} {\cal F}(x,k_\perp^2) \sim \sum_N \bar{\alpha}^N \cdot \frac{l^N}{N!} \cdot \frac{\kappa^N}{N!} = I_0(2\sqrt{\bar{\alpha} l \kappa})
\label{DGLAP}
\end{equation}
  
For smaller $k_\perp$, in the BFKL region, also non-ordered chains contribute, and the result is a powerlike increase $\sim x^{-\lambda}$ for small $x$-values \cite{BFKL}.

The CCFM \cite{CCFM} model interpolates smoothly between the DGLAP and BFKL regimes. The Linked Dipole Chain (LDC) model \cite{LDC,jimhamid,hamid} is a reformulation and generalization of the CCFM model, in which gluons in some parts of phase space are treated as final state radiation, instead of as initial state radiation as in the CCFM model. In the LDC model the  possibility to ``go down'' in $k_\perp$, from $\kappa'$ to a smaller value $\kappa$, is suppressed by a factor $\exp(\kappa - \kappa')$. The effective allowed distance for downward steps is therefore limited to about one unit in $\kappa$. If this quantity is called $\delta$, the result is that the phase space factor $\frac{\kappa^N}{N!}$ is replaced by $\frac{(\kappa+N\delta)^N}{N!}$. Thus we obtain
 
\begin{equation}
\frac{1}{\bar{\alpha}} {\cal F}(x,k_\perp^2) \sim \sum_N \frac{(\bar{\alpha} l)^N}{N!} \frac{(\kappa+N\delta)^N}{N!}
\label{fixsumma}
\end{equation}
When $\kappa$ is very large, this expression approaches eq. (\ref{DGLAP}). When $\kappa$ is small we find, using Sterling's formula, that ${\cal F}(x,k_\perp^2) \sim \sum_N (\bar{\alpha} l \delta e)^N / N! = \exp (\lambda l) = x^{-\lambda}$, with $\lambda = \bar{\alpha} \delta e$. 
For $\delta = 1$ this gives $\lambda = e \bar{\alpha} = 2.72\bar{\alpha}$, which is very close to the leading order result for the BFKL equation, $\lambda = 4 \ln 2 \, \bar{\alpha} = 2.77\bar{\alpha}$.
Thus eq. (\ref{fixsumma}) interpolates smoothly between the DGLAP and BFKL regimes. It is also possible to see that the transition occurs for a fixed ratio between $\kappa$ and $l$, $\kappa / l \approx e \bar{\alpha}$.

For small $x$ we can also go one step further, and expand eq.~(\ref{fixsumma}) in powers of $\kappa/l$. After some algebra we then find
\begin{equation}
\frac{1}{\bar{\alpha}} {\cal F} \sim \frac{1}{\sqrt{2 \pi \lambda l}} \exp \left[\lambda l + \frac{\kappa}{\delta} - \frac{\kappa^2}{2 \delta^2 \lambda l}\right].
\label{Gauss}
\end{equation}
We here recognize the Gaussian distribution with a width proportional to $\lambda l$, corresponding to a random walk in $\kappa$, typical for the BFKL equation \cite{BFKL,cigar}. 

\begin{figure}
\mbox{\epsfig{file=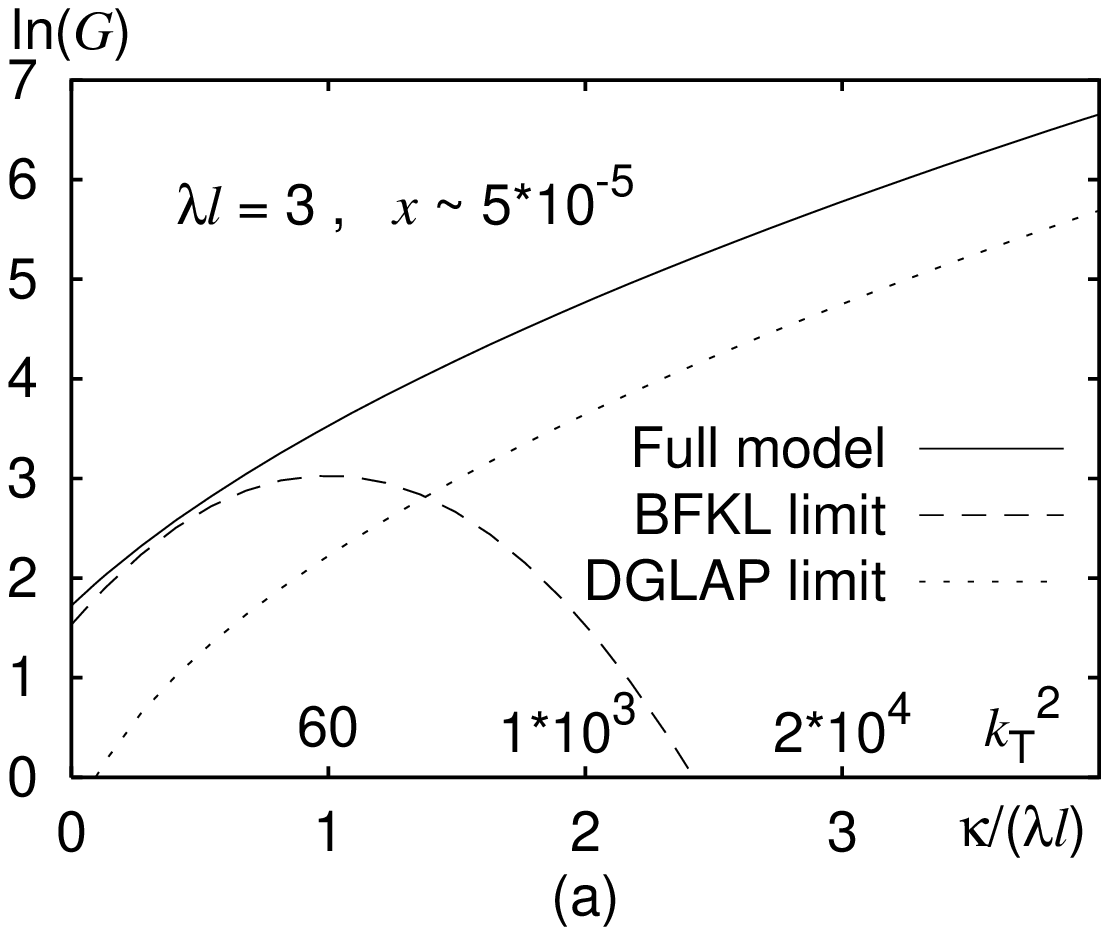,width=6.1cm}}
\mbox{\epsfig{file=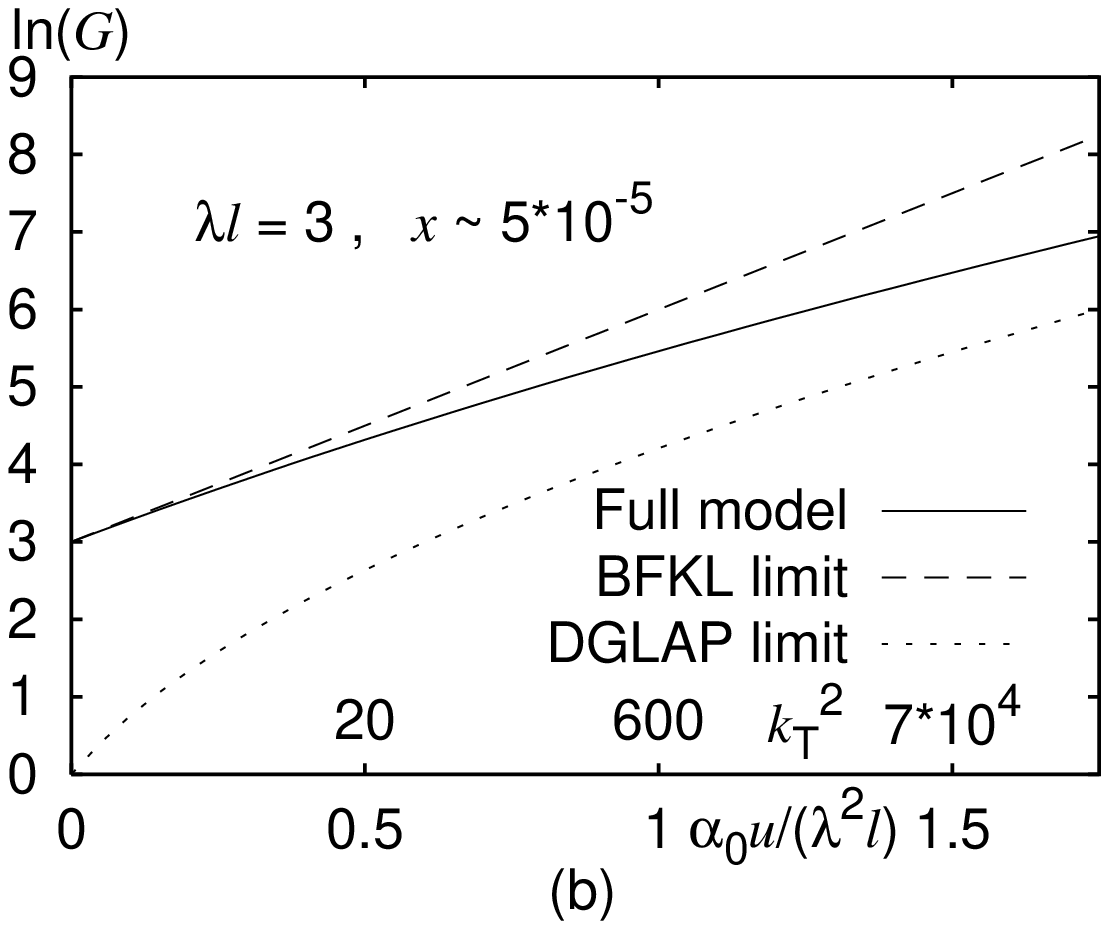,width=6.1cm}}
\caption{\label{figure}\em The structure function $G= (1/\bar{\alpha}) {\cal F}$  for fixed coupling, a, and running coupling, b. The combinations on the abscissa, $\kappa/(\lambda l)$ and $\alpha_0 u/(\lambda^2 l)$, are chosen so that the transition occurs when these combinations are around 1. The corresponding values of $k_\perp^2$ in $GeV^2$ are also indicated.}
\end{figure}

The result of a numerical evaluation of eq.~(\ref{fixsumma}) is presented in fig.~\ref{figure}a together with the DGLAP and BFKL approximations in eqs.~(\ref{DGLAP}) and (\ref{Gauss}). 
We note in particular that ${\cal F}$ grows monotonically with $\kappa$. The Gaussian behaviour is only obtained in an expansion of $\ln {\cal F}$ to second order in powers of $\kappa/l$.

\section{Running Coupling}

For a running coupling a factor $\bar{\alpha}\, d\kappa$ in eq. (\ref{dglap}) is replaced by $\alpha_0\, d \kappa/\kappa = \alpha_0\, du$, where $\bar{\alpha} \equiv \alpha_0 / \kappa$ and $u\equiv \ln \kappa=\ln\ln k_\perp^2$. 
In the large $k_\perp$ region (the ``DGLAP region'') the result is therefore similar to eq. (\ref{DGLAP}), but with $\bar{\alpha} \kappa$ replaced by $\alpha_0 \ln \kappa$.
For small $x$-values we note that the allowed effective size of downward steps, which is still one unit in $\kappa$, is a sizeable interval in $u$ for small $\kappa$, but a very small interval in $u$ for larger $\kappa$. 
This is the reason for  our earlier experience \cite{jimhamid}, which showed  that a typical chain contains two parts. In the first part the $k_\perp$-values are relatively small, and it is therefore easy to go up and down in $k_{\perp}$, and non-ordered $k_\perp$-values are important. The second part is an ordered (DGLAP-type) chain, where $k_\perp$ increases towards its final value. 

Let us study a chain with $N$ links, out of which $N-k$ correspond to the first part with small non-ordered $k_\perp$, and the remaining $k$ links belong to the second part with increasing $k_\perp$. Assume that the effective space for each $u_i$ in the soft part is given by a quantity $\Delta$. ($\Delta$ is sensitive to the soft $k_\perp$-cutoff, and an estimate is presented in ref. 1.) Then the total weight for this part becomes $\Delta^{N-k}$. For the $k$ links in the second, ordered, part the phase space becomes as in the DGLAP case $u^k/k!$. Thus the total result is ($G \equiv {\cal F}/\bar{\alpha} = \kappa {\cal F}/\alpha_0$)
\begin{equation}
G \sim \sum_N \frac{(\alpha_0 l)^N}{N!} \sum_{k=0}^N \frac{u^k}{k!} \Delta^{N-k} 
=\sum_N \frac{(\alpha_0 l \Delta)^N}{N!} \sum_{k=0}^N \frac{(u/\Delta)^k}{k!}\,\,;\,\, u \equiv \ln \kappa.
\label{Gslut}
\end{equation}

This simple model also interpolates smoothly between the DGLAP and BFKL regions. For large $u$-values the last term in the sum over $k$ dominates, which gives the ``DGLAP'' result
\begin{equation}
G \sim \sum (\alpha_0 l u)^N/(N!)^2 = I_0(2\sqrt{\alpha_0 l u}) \approx (16 \pi^2 \alpha_0 l u)^{-1/4} \cdot \exp(2\sqrt{\alpha_0 l u}).
\label{dglaprunning}
\end{equation}
For small $u$-values the sum over $k$ gives approximately $\exp u/\Delta$, and thus
\begin{equation}
G \sim \exp (\alpha_0 \Delta l) \cdot \exp(u/\Delta) =x^{-\lambda} \kappa^{\lambda/\alpha_0}, \,\,\mathrm{with}\,\,\, \lambda=\alpha_0 \Delta.
\label{bfklrunning}
\end{equation}
The transition between the regimes occurs now for a fixed ratio between  $u=\ln \ln k_\perp^2$ and $l=\ln(1/x)$,
$u / l  \approx \lambda ^2 / \alpha_0$, which is of order 0.1 if $\lambda$ is around 0.3. 

The result of a numerical evaluation is illustrated in fig. \ref{figure}b. The DGLAP and BFKL limits in eqs. (\ref{dglaprunning}) and (\ref{bfklrunning}) are also presented, and we can see how the model interpolates between these limits.

\section{Conclusion}
The simple models in eqs. (\ref{fixsumma}) and (\ref{Gslut}), for fixed and running couplings respectively, interpolate smoothly between large and small $k_\perp$-values. They contain the essential features of the CCFM and LDC models, and can therefore give an intuitive picture of the transition between these two regimes. The transition occurs for a fixed ratio between $\ln k_\perp^2$ and $\ln 1/x$ for fixed coupling, and between $\ln  \ln k_\perp^2$ and $\ln 1/x$ for a running coupling. A more extensive discussion, including effects of non-leading terms in $\ln 1/x$ and the properties of Laplace transforms, can be found in ref. 1.


\begin{thebibliography}{99}

\bibitem{paper}
G. Gustafson and G. Miu, hep-ph/0110143

\bibitem{DGLAP}
V.N. Gribov and L.N. Lipatov, Sov. J. Nucl. Phys. {\bf 15} (1972) 438 and 675; G. Altarelli and G. Parisi, Nucl. Phys. {\bf B126} (1977) 298; Yu.L. Dokshitzer, Sov. Phys. JETP {\bf 46} (1977) 641

\bibitem{BFKL}
E.A. Kuraev, L.N. Lipatov and V.S. Fadin, Sov. Phys. JETP {\bf 45} (1977) 199; Ya.Ya. Balitsky and L.N. Lipatov, 
Sov. J. Nucl. Phys. {\bf 28} (1978) 822

\bibitem{CCFM}
M. Ciafaloni, Nucl. Phys. {\bf B296} (1988) 49;
S. Catani, F. Fiorani and G. Marchesini, Phys. Lett. 
{\bf B234} (1990) 339, 
Nucl. Phys. {\bf B336} (1990) 18.

\bibitem{LDC}
B. Andersson, G. Gustafson and J. Samuelsson, Nucl. Phys. 
{\bf B467} (1996) 443;
H. Kharraziha and L. L\"onnblad, JHEP {\bf 03} (1998) 006

\bibitem{jimhamid}
B. Andersson, G. Gustafson, H. Kharraziha and J. Samuelsson, Z. Phys. 
{\bf C71} (1996) 613

\bibitem{hamid}
B. Andersson, G. Gustafson and H. Kharraziha, Phys.Rev. 
{\bf D57} (1998) 5543

\bibitem{cigar}
E.M. Levin and M.G. Ryskin, Phys. Rep. {\bf 189} (1990) 267;
J. Bartels and H. Lotter, Phys. Lett. {\bf B309} (1993) 400;
J.R. Forshaw, P.N. Harriman and P.J. Sutton, Nucl. Phys. {\bf B146} (1994) 739

\end{thebibliography}
\end{document}